\documentclass[11pt]{article}
\usepackage[utf8]{inputenc}

\usepackage[margin=1in]{geometry}

\usepackage{xcolor}
\title{Draft notes}
\date{}

\usepackage[utf8]{inputenc}                      
\usepackage{subfig}
\usepackage[T1]{fontenc} 
\usepackage{graphicx}
\usepackage{epsfig}
\usepackage{amsmath}
\usepackage{amssymb}
\usepackage{float}
\usepackage{placeins}
\usepackage[utf8]{inputenc}
\usepackage{changepage}
\usepackage{bm}
\usepackage{multicol}
\usepackage{comment}
\usepackage{shuffle}
\usepackage[font=small,labelfont=bf]{caption}

\allowdisplaybreaks[4]

\DeclareMathOperator*{\Reglim}{Reg}

\newcommand{\Hf}[5]{ 
 {}_{#1}F_{#2}  \left (\begin{smallmatrix}
   {#3} \\
   {#4}
\end{smallmatrix}\Bigr|#5\right)
  }

\newcommand{\Ef}[3]{ 
    E_4  \left (\begin{smallmatrix}
   {#1} \\
   {#2}
\end{smallmatrix};#3\right)
  }

\newcommand{\Eff}[5]{ 
    E_4  \left (\begin{smallmatrix}
   {#1} & {#3} \\
   {#2} & {#4}
\end{smallmatrix};#5\right)}

\newcommand{\Efff}[7]{ 
    E_4  \left (\begin{smallmatrix}
   {#1} & {#3} & {#5}\\
   {#2} & {#4} & {#6}
\end{smallmatrix};#7\right)}

\newcommand{\z}{&&\hspace*{-1cm}}

\newcommand{\ep}{\varepsilon}

\newcommand{\bea}{\begin{eqnarray}}
\newcommand{\eea}{\end{eqnarray}}
\newcommand{\be}{\begin{equation}}
\newcommand{\ee}{\end{equation}}

\begin{document}

\begin{titlepage}
\renewcommand{\thefootnote}{\fnsymbol{footnote}}

\begin{center}
{\Large \bf Sunrise integral in non-relativistic QCD with elliptics}
\end{center}

\par \vspace{2mm}
\begin{center}
{\sc A.~Kotikov}
  \vspace{5mm}

  {\normalsize\it Bogolyubov Laboratory for Theoretical Physics, JINR,}\\
{\normalsize\it 141980 Dubna (Moscow region), Russia} \\

\vspace{5mm}

\end{center}

\par \vspace{2mm}
\begin{center} {\large \bf Abstract} \end{center}
\begin{quote}
\pretolerance 10000

The main steps of the process of obtaining the result \cite{Campert:2020yur} in terms of elliptic polylogarithms for a two-loop sunrise integral with two different
  internal masses with pseudothreshold kinematics for all orders of the dimensional regulator are shown.

\end{quote}

\vspace*{\fill}

\end{titlepage}

\tableofcontents

\section{Introduction}
\label{sec:intro}

Feynman integrals allow a Laurent expansion with respect to a dimensional regulator, and the coefficients of this expansion can often be explicitly computed in terms
of well-known special functions such as multiple polylogarithms (MPL) and multiple elliptic polylogarithms (eMPL) (see the recent paper \cite{Bourjaily:2022bwx}
and references and discussions therein). In practice, it is often possible to truncate the
Laurent series, since only a few orders of expansion are required to calculate physically significant quantities. Nevertheless, it is interesting to investigate the
analytical structure of these coefficients at higher orders, or, more generally, at all orders of the dimensional regulator.

This article shows the main stages of the consideration  \cite{Campert:2020yur} of a two-loop sunrise integral topology with two different
internal masses, denoted $m$ and $M$, and pseudo-threshold kinematics $p^2=-m^2$
\cite{Kalmykov:2008ge} (see also \cite{Kniehl:2005bc,Kniehl:2019vwr}).
This integral family arises when considering the nonrelativistic limit of Quantum Chromodynamics.

The analytical structure of the sunrise topology considered in  \cite{Campert:2020yur} was studied using differential equations in
\cite{Kotikov:1990kg,Kotikov:1991hm,Kotikov:1991pm,Bern:1993kr,Remiddi:1997ny} and using the effective mass analysis in
\cite {Kotikov:1990zs,Kniehl:2012hn,Kotikov:2020ccc} (see the recent review in \cite{Kotikov:2021tai}). Moreover, this integral family admits a closed-form solution in terms of ${}_3F_{2}$-hypergeometric functions, as shown
in \cite{Kalmykov:2008ge} (the corresponding off-shell diagrams with equal masses are much more complicated and their explicit solution requires hypergeometric
Appell functions $F_2$ \cite{Tarasov:2006nk}). In Ref.  \cite{Campert:2020yur}, we
obtained an expression in terms of eMPL, valid for all orders of the dimensional regulator.\footnote{Similar results in more
  complicated cases can be found in Refs. \cite{Adams:2017ejb,Bezuglov:2020ywm,Bezuglov:2021tax,Bezuglov:2022npo}.}

\label{sec:preparation}
\section{The sunrise integral}

Following Ref. \cite{Kalmykov:2008ge} we
study the sunrise integral topology defined as,
\begin{equation}
    J_{i_1,i_2,i_3}(m^2,M^2)=\left.\int\int \frac{d^D k_1 d^D k_2}{[k_2^2-m^2]^{i_1}[k_1^2-M^2]^{i_2}[(k_1-k_2-q)^2-M^2]^{i_3}}\right|_{q^2=-m^2},
\end{equation}
with $D=4-2\epsilon$. As it was observed in  \cite{Campert:2020yur},
\be
J_{1,2,2} = \hat{N}_1 \biggl[J^{(1)}_{1,2,2}(t) - (2t)^{\ep-1} \, J^{(2)}_{1,2,2}(t) - (2t)^{\ep} \, J^{(3)}_{1,2,2}(t) \biggr],~~\left(t=\frac{m^2}{2 M^2}\right) \, ,
\label{J122N}
\ee
where,
\bea
\z J^{(1)}_{1,2,2}(t) = \frac{1+\ep}{6\ep (1-\ep)} \, \Hf{4}{3}{1,\frac{3}{2},1+\frac{\epsilon}{2},\frac{3}{2}+\frac{\epsilon}{2}}{2-\epsilon,\frac{5}{4},\frac{7}{4}}{-t^2} = - \frac{\hat{K}}{2^{2\epsilon+2}\ep t^2} \, I^{(1)}(t)\,, \nonumber \\
\z J^{(2)}_{1,2,2}(t) = \frac{1}{(1+2\ep)(1-\ep)}\,  \Hf{4}{3}{1,\frac{1}{2}+\epsilon,1+\frac{\epsilon}{2},1+\epsilon}{\frac{3}{2}-\frac{\epsilon}{2},\frac{3}{4}+\frac{\epsilon}{2},\frac{5}{4}+\frac{\epsilon}{2}}{-t^2} =  \frac{\hat{K}}{2^{4\epsilon+1} t^{1-\ep}} \, I^{(2)}(t)\, , \nonumber \\
\z J^{(3)}_{1,2,2}(t) = \frac{1+\ep}{\ep (2-\ep)(3+2\ep)} \,\Hf{4}{3}{1,\frac{3}{2}+\frac{\epsilon}{2},1+\epsilon,\frac{3}{2}+\epsilon}{2-\frac{\epsilon}{2},\frac{5}{4}+\frac{\epsilon}{2},\frac{7}{4}+\frac{\epsilon}{2}}{-t^2} = - \frac{\hat{K}}{2^{4\epsilon+2}\ep t^2} \, I^{(3)}(t)
\label{Ji122N1}
\eea
and $\hat{K}$ is defined as,
\begin{equation}
\hat{K}=\frac{\Gamma(1-\ep)}{\Gamma(1-2\ep)\Gamma(1+\ep)},
\end{equation}
while the factors $I^{(i)}(t)$ represent the relevant
integrals,
\begin{align}
\label{eq:2integrals p}
     I^{(1)}(t)&=\int_0^1 \,dp \,p^{\epsilon -1}(1-p)^{-\epsilon -\frac{1}{2}} \left((p^2 t^2+1)^{-\frac{1}{2}}(1-\ep J_1(p)) -1\right)  ,\nonumber\\
     I^{(2)}(t)&=  \int_0^1 \,dp\, p^{3 \epsilon -1} (1-p)^{-\epsilon -\frac{1}{2}}  \left(p^2 t^2+1\right)^{-\epsilon -\frac{1}{2}}\, J_2(p),
     \nonumber\\
     I^{(3)}(t)&=\int_0^1\, dp \,p^{2 \epsilon -1}(1-p)^{-\epsilon -\frac{1}{2}} \left(\left(p^2 t^2+1\right)^{-\frac{\epsilon}{2} -\frac{1}{2}}(1-\frac{\ep}{2} J_3(p))-1\right),
\end{align}
with
\bea
&&J_1(p)=q(p)^{\epsilon }\int_0^{q(p)} dz\left((1-z)^{-\frac{1}{2}}-1\right) z^{-\epsilon -1},~~ q(p)=\frac{p^2 t^2}{p^2 t^2+1},\nonumber\\
&&J_2(p)=\int_{0}^{pt} dz \,z^{-\epsilon } \left(z^2+1\right)^{\epsilon -\frac{1}{2}},\nonumber\\
&&J_3(p)=q(p)^{\frac{\epsilon}{2}}\int_0^{q(p)} dz \left((1-z)^{-\frac{\epsilon}{2} -\frac{1}{2}}-1\right) z^{-\frac{\epsilon }{2}-1},
\label{Ji}
\eea

We remark that integral $J_{1,2,2}$ has a finite $\epsilon$ expansion. Other sunrise integrals, $J_{1,1,2}$ and $J_{1,1,1}$, considered in Ref.  \cite{Campert:2020yur},
contain singularities but
their consideration is beyond the slope of this short paper.

\section{All orders result in terms of elliptic polylogarithms}
\label{sec:all orders empls}

We are interested in the computation of iterated integrals of the form,
\begin{equation}
\label{eq:GenericItIntEll}
    \int_0^{x} dx_1 R_1(x_1,y(x_1)) \int_0^{x_1} dx_2 R_2(x_2,y(x_2))\dots\int_0^{x_{n-1}} dx_n R_{n}(x_n,y(x_n))\, ,
\end{equation}
where $R_i$ are rational functions of their arguments and $y(x)$ is an elliptic curve,
\begin{equation}
y(x)=\sqrt{(x-a_1)(x-a_2)(x-a_3)(x-a_4)}\,,
\end{equation}
All iterated integrals of the form (\ref{eq:GenericItIntEll}) can be expressed in terms of eMPLs:
\begin{equation}
\label{eq:E4_def}
\Ef{n_1,  \dots,  n_k}{c_1,  \dots, c_k}{x} = \int_0^xdt\,\varphi_{n_1}(c_1,t)\,\Ef{n_2,  \dots , n_k}{c_2,  \dots, c_k}{t},~~\Ef{}{}{x}=1\,.
\end{equation}
By construction, the kernels $\varphi_n(c,x)$ have at most simple poles, and they are (see \cite{Broedel:2017kkb} for a detailed discussion)
\begin{align}
\label{PhiFunc}
& \varphi_0(0,x)= \frac{c_4}{y(x)}\,,~~
\varphi_1(c,x)= \frac{1}{x-c}\,,~~ \varphi_{-1}(\infty,x) = \frac{x}{y(x)}\,, \nonumber\\&
\varphi_{-1}(c,x) = \frac{y(c)}{(x-c)y(x)}-(\delta_{c0}+\delta_{c1})\frac{1}{x-c}, ...
\end{align}
where
\begin{equation}
 c_{4} = \frac{1}{2}\sqrt{a_{13}a_{24}} \quad \text{with} \quad a_{ij}=a_i-a_j\, .
\end{equation}   
Moreover we define,
\begin{equation}
    \Ef{\vec{1}}{\vec{0}}{x}\equiv\frac{\log(x)^n}{n!},
\end{equation}
where $\vec{1}$ and $\vec{0}$ are vectors with elements 
equal to $1$ and $0$ respectively, and $n=\text{length}(\vec{1})=\text{length}(\vec{0})$.

EMPLs
are a generalisation of MPLs
defined recursively as,
\begin{equation}
    G(a_1,a_2,\ldots,a_n;x)=\int_0^x \frac{dt}{t-a_1} G(a_2,\ldots,a_n,t),~~G(;x)\equiv 1\,,
\end{equation}
and,
\begin{equation}
    G(\vec{0},x)\equiv\frac{\log(x)^n}{n!}\,.
\end{equation}
By definition we see that MPLs are a subset of eMPLs,
\begin{equation}
    \Ef{1,  \dots,  1}{c_1,  \dots, c_n}{x} = G(c_1,c_2,\dots,c_n;x),~~c_i \neq \infty \,.
\end{equation}

As with all iterated integrals, eMPLs satisfy a shuffle algebra with the shuffle product defined as
\begin{equation}
 \Efff{a_1}{a'_1}{\dots}{\dots}{a_n}{a'_n}{x}\Efff{b_1}{b'_1}{\dots}{\dots}{b_m}{b'_m}{x} =  \sum_{\vec{c}=\vec{a}\shuffle\vec{b}} \Efff{c_1}{c'_1}{\dots}{\dots}{c_{n+m}}{c'_{n+m}}{x}\, .
\end{equation}
The vector $\vec{c}$ is the vector obtained by shuffling all $\vec{a}$ and $\vec{b}$ while preserving the order of the elements $\vec{a}$ and $\vec{b}$ respectively.

{\bf Regularisation}~~
\label{sec:regularisation}
As we will see below,
we are interested in computing definite integrals of the form,
\begin{equation}
\label{eq:reg_ex}
    \int_0^1 f(x)dx= F(1)-F(0) , \quad \frac{\partial F(x)}{\partial x}=f(x)\,.
\end{equation}
In some cases, the primitive is ill-defined when evaluated on the boundaries of integration, and two constraints must be satisfied in order to evaluate the definite integral:
\begin{equation}
\label{eq:reg_exlim_0}
    \int_0^1 f(x)dx= \lim_{x\rightarrow 1}F(x)-\lim_{x\rightarrow 0}F(x) \equiv \Reglim{}_{0,1} F(x).
\end{equation}

{\bf Compact representation.}
The analysis in \cite{Campert:2020yur}  implies that all integrals
are formally evaluated as
\begin{equation}
\label{eq: gen eMPLs 2-integral}
\sum_{l=1}^{n'} C_l\sum_{i,j=0}^\infty \frac{\epsilon^{i+j}}{i!j!}\int_0^1 dx k_{1,l}(x) L^i_{1,l}(x)\int_1^x dz k_{2,l}(z) L^j_{2,l}(z),
\end{equation}
where $C_l$ are some coefficients,
$L_{i,j}(x)$ are
combinations of eMPLs of depth one, while $k_{i,j}(x)$ are
combinations of integration kernels.
These integrals can be directly computed in terms of eMPL by shuffle expanding the eMPL products of the integrands and recursively using the definition of eMPL.

To make the notation more compact and to make the properties of the result in terms of eMPL clear, we use the following notation for double integrals of
(\ref{eq: gen eMPLs 2-integral}). Denoting the primitive $k_{i,j}(x)$ as $K_{i,j}(x)$ and defining the bilinear $*$-operator as,
\begin{equation}
\Ef{\vec{n}}{\vec{c}}{x}*  \Ef{\vec{m}}{\vec{d}}{x}=    \Eff{\vec{n}}{\vec{c}}{\vec{m}}{\vec{d}}{x},
\label{star}
\end{equation}
we can write (\ref{eq: gen eMPLs 2-integral}) in the following form
\begin{equation}
\label{eq: gen eMPLs * notation}
   \Reglim_{0,1} \sum_{l=1}^{n'} C_l\sum_{i,j=0}^\infty \frac{\epsilon^{i+j}}{i!j!} K_{1,l}(x) * L_{1,l}(x)^i \left[ K_{2,l}(z) * L_{2,l}(z)^j\right]_1^x,
\end{equation}
where all eMPL products are shuffle expanded before the $*$ operator is applied, and these operations are performed on the inner square brackets first. Finally, the
lower and upper scripts applied to the square brackets indicate the following operation:
\begin{equation}
    \left[F(x)\right]_1^x= F(x)-F(1)\,.
\end{equation}

\section{The integral $I^{(2)}(t)$ }
We show how the solution strategy of the previous section works in practice by considering the integral $I^{(2)}(t)$ in Eq. (\ref{eq:2integrals p}).
The dependence on the elliptic curve is made explicit by changing the variable $p \to (1-x^2)$, 
\begin{equation}
\label{eq:I2}
I^{(2)}(t)=\int_0^1 dx \;\frac{2}{t \left(1-x^2\right) y(x)}\left(\frac{\left(1-x^2\right)^3}{t^2 x^2 y(x)^2}\right)^{\epsilon } 
   \int_0^{t(1- x^2)}dz \frac{1}{\sqrt{z^2+1}} \left(z+\frac{1}{z}\right)^{\epsilon },
\end{equation}
where the inner integral can be expressed as
\begin{equation}
   \int_0^{t-t x^2}dz \frac{1}{\sqrt{z^2+1}} \left(z+\frac{1}{z}\right)^{\epsilon }=-\int_1^x dz \frac{2 z}{y(z)} \left(\frac{t y^2(z)}{1- z^2}\right)^{\epsilon } .
\end{equation}
All the $\varepsilon$-powers are
expanded in $\varepsilon$:
\begin{equation}
\left(\frac{t y^2(x)}{1- x^2}\right)^{\epsilon }=\sum_{i=0}^\infty\frac{\epsilon ^i }{i!}\log ^i\left(\frac{t y^2(x)}{1- x^2}\right)\,.
\end{equation}
The resulting logarithm is
expressed in terms of eMPLs as
\begin{equation}
  \log\left(\frac{t y^2(x)}{1- x^2}\right)= \log\bigl(t y^2(0)\bigr)
  + \int_0^x dz \frac{2 z \left(t^2 \left(z^2-1\right)^2-1\right)}{t^2 \left(z^2-1\right) y(z)^2}\,,
\end{equation}
where $y^2(0)=t^{-2}+1$.
The integrand can be written in terms of the integration kernels:
\begin{equation}
    \frac{2 z \left(t^2 \left(z^2-1\right)^2-1\right)}{t^2 \left(z^2-1\right) y(z)^2}=\sum_{i=1}^4\varphi _1\left(a_i,z\right)-\varphi _1(-1,z)-\varphi _1(1,z)\,,
\end{equation}
where we denoted with $a_i$ the four roots of the elliptic curve,
\begin{equation}
    a_1=-\frac{\sqrt{t-i}}{\sqrt{t}}\,,\; a_2=\frac{\sqrt{t-i}}{\sqrt{t}}\,, \; a_3=-\frac{\sqrt{t+i}}{\sqrt{t}}\,, \; a_4=\frac{\sqrt{t+i}}{\sqrt{t}}\,.
\end{equation}
Upon integration we find,
\begin{equation}
 L_4\equiv \log\left(\frac{t y^2(x)}{1- x^2}\right)=  \sum_{i=1}^4\Ef{1}{a_i}{x}-\Ef{1}{-1}{x}-\Ef{1}{1}{x}+\log \left(t^2+1\right)-\log (t)\,.
\end{equation}
Finally, all prefactors can be expressed in terms of integration kernels
\begin{equation}
\label{eq:Ker}
    \frac{2}{t \left(1-x^2\right) y(x)}=\varphi _{-1}(-1,x)-\varphi _{-1}(1,x)-\varphi _1(1,x)\,.
\end{equation}
By taking the primitive of Eq.~(\ref{eq:Ker}) we obtain,
\begin{equation}
    K_4\equiv\Ef{-1}{-1}{x}-\Ef{-1}{1}{x}-\Ef{1}{1}{x}.
\end{equation}

Applying these methods to all relevant logarithms and prefactors, we get the result in terms of integrals of the form (\ref{eq: gen eMPLs 2-integral}), which are directly evaluated in eMPL, for example, by Eq.~(\ref{eq: gen eMPLs * notation}):
\begin{equation}
   I^{(2)}(t)=\Reglim_{0,1}\sum_{i,j=0}^\infty\frac{\epsilon ^{i+j}}{i!j!}K_4*  L_5^{i}  \left[K_5 * L_4^{j}\right]_1^x\,,
\end{equation}
where,
\begin{align}
    L_5&=-\sum_{i=1}^4\Ef{1}{a_i}{x}+3 \Ef{1}{-1}{x}-2 \Ef{1}{0}{x}+3 \Ef{1}{1}{x}-\log \left(t^2+1\right), \nonumber \\
    K_5&=-2 \Ef{-1}{\infty }{x}.
\end{align}

\section{Conclusions}
\label{sec:conclusions}
In this short paper, we have shown the main stages of the study  \cite{Campert:2020yur} of the sunrise integral $J_{1,2,2}$ with two different internal masses and
pseudothreshold kinematics in dimensional regularization. This integral admits a closed form solution in terms of hypergeometric functions \cite{Kalmykov:2008ge}, and we have used this representation
as the starting point of our analysis. In particular, in  \cite{Campert:2020yur} we shown that all relevant hypergeometric functions can be represented as iterated integrals depending on
one elliptic curve
(see also Eq. (\ref{eq:2integrals p})). When these integrals are expanded in terms of a
dimensional regulator, the expansion coefficients are iterated integrals in terms of rational functions on the corresponding elliptic curve with at most simple poles.
Calculating $I^{(2)}_1(t)$ in Eq. (\ref{eq:2integrals p}), we have shown a way to represent the sunrise integral $J_{1,2,2}$ in terms of eMPLs, which is true for all orders
of the dimensional regulator.\\

{\bf Acknowledgments:} Author thanks the Organizing Committee of the International Conference on Quantum Field Theory, High-Energy Physics, and Cosmology
 for the invitation.

\bibliographystyle{unsrt}

\bibliography{EllipK.bib}

\end{document}